\documentclass[sigconf] {acmart}
\usepackage{stfloats}
\usepackage{graphbox}
\usepackage{float}
\usepackage{multicol}
\usepackage{subcaption}
\usepackage{xcolor}
\usepackage{placeins}
\usepackage{afterpage}
\usepackage{caption}

\AtBeginDocument{%
  }

\copyrightyear{2025}
\acmYear{2025}
\setcopyright{rightsretained}
\acmConference[CHI EA '25]{Extended Abstracts of the CHI Conference on
Human Factors in Computing Systems}{April 26-May 1, 2025}{Yokohama, Japan}
\acmBooktitle{Extended Abstracts of the CHI Conference on Human Factors in
Computing Systems (CHI EA '25), April 26-May 1, 2025, Yokohama,
Japan}\acmDOI{10.1145/3706599.3719806}
\acmISBN{979-8-4007-1395-8/2025/04}

\begin{document}
\title{Cybersafety Card Game: Empowering Digital Educators to Teach Cybersafety to Older Adults}

\author{Jacob Camilleri}
\orcid{0009-0009-4912-258X}
\affiliation{%
  \institution{Munster Technological University}
  \city{Cork}
  \country{Ireland}
}
\email{jacob.camilleri@mtu.ie}

\author{Ashley Sheil}
\orcid{0000-0001-7750-9495}
\affiliation{%
  \institution{Munster Technological University}
  \city{Cork}
  \country{Ireland}
  }  
\email{ashley.sheil@mtu.ie}

\author{Michelle O Keeffe}
\orcid{0009-0008-7130-9941}
\affiliation{%
  \institution{Munster Technological University}
  \city{Cork}
  \country{Ireland}
}
\email{michelle.okeeffe2@mtu.ie}

\author{Moya Cronin}
\orcid{0009-0008-2831-186X}
\affiliation{%
  \institution{Munster Technological University}
  \city{Cork}
  \country{Ireland}
}
\email{moya.cronin@mycit.ie}

\author{Melanie Gruben}
\orcid{0009-0008-9332-5205}
\affiliation{%
  \institution{Munster Technological University}
  \city{Cork}
  \country{Ireland}
}
\email{melanie.gruben@mtu.ie}

\author{Hazel Murray}
\orcid{0000-0002-5349-4011}
\affiliation{%
 \institution{Munster Technological University}
 \city{Cork}
 \country{Ireland}}
 \email{hazel.murray@mtu.ie}

\renewcommand{\shortauthors}{Camilleri et al.}

\begin{abstract} 
Digital inequality remains a significant barrier for many older adults, limiting their ability to navigate online spaces securely and confidently while increasing their susceptibility to cyber threats.
In response, we propose a novel shedding-type card game for older adults to conceptually learn and reinforce cyber hygiene practices in educational settings.
We asked digital educators to participate as players alongside older adults ($n = 16$), departing from their usual role as teachers, 
they collaborated and shared a unique learning experience. 
The cybersafety game addresses 4 key topics: handling scams, password management, responding to cyber attacks, and staying private. We adopted a mixed-method approach of think-aloud playtesting, semi-structured interviews, and surveys to evaluate the game's reception and impact. Participants reported highly favorable gameplay experiences and found the cybersafety advice useful. Player feedback informed game modifications, detailed in this paper, to further enhance the game's usability and educational value. 

\end{abstract}



\maketitle

\begin{figure*}
  \centering 
  \subfloat[]{\includegraphics[width=0.2\textwidth]{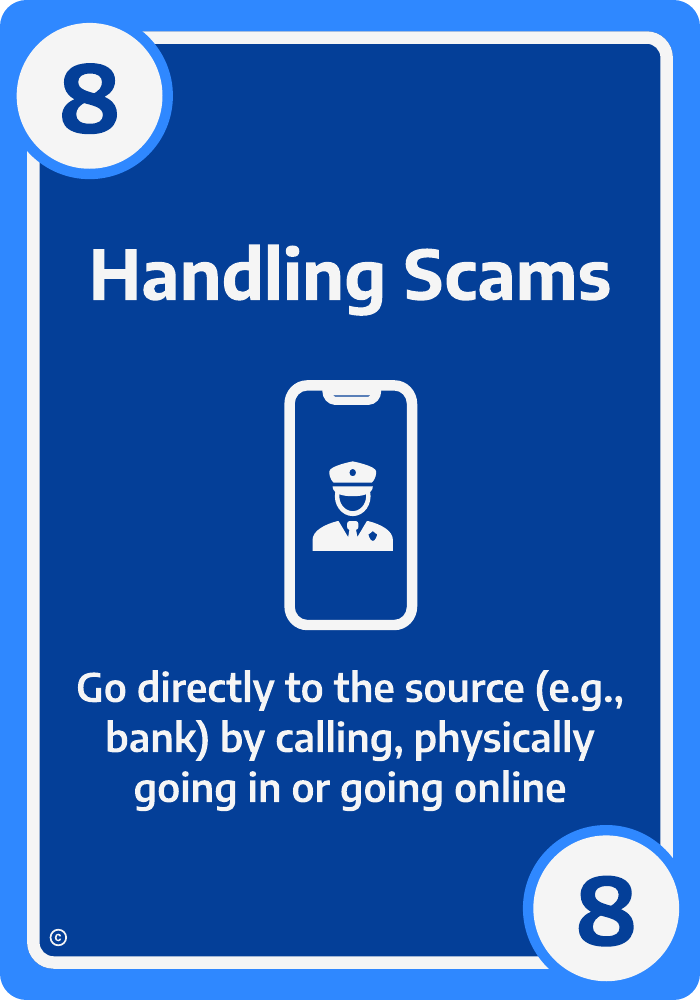}} 
  \label{HS}
  \subfloat[]{\includegraphics[width=0.2\textwidth]{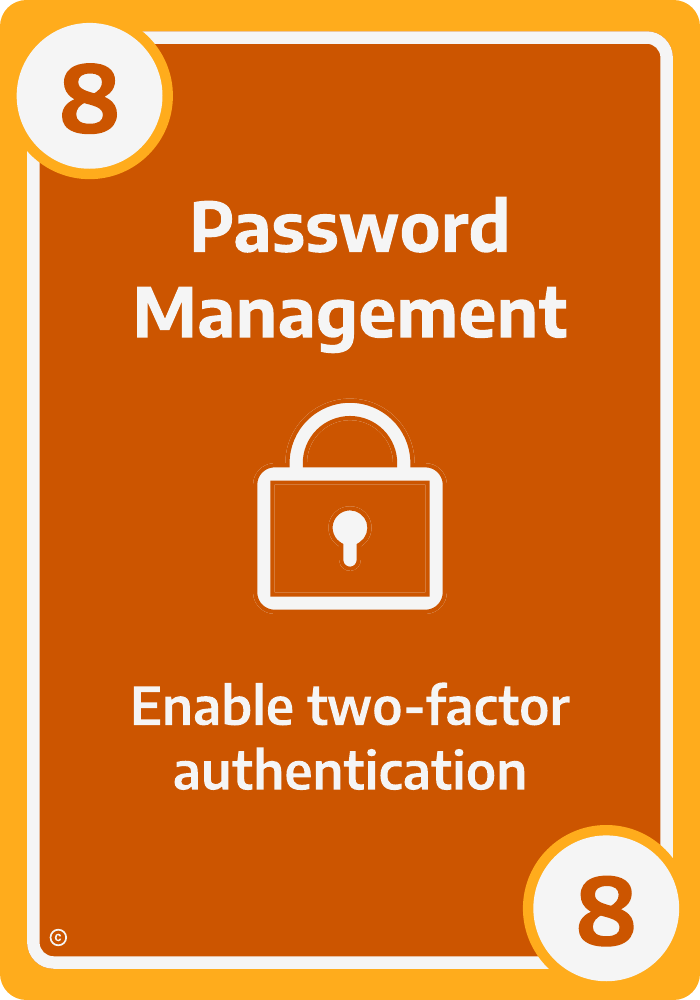}} 
  \subfloat[]{\includegraphics[width=0.2\textwidth]{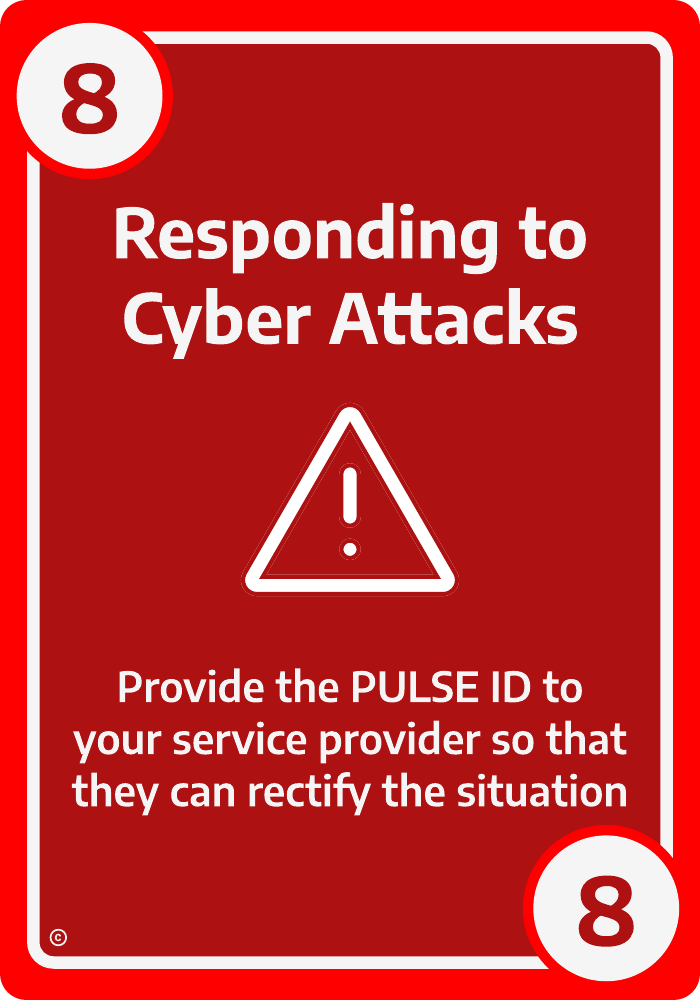}}\\
\subfloat[]{\includegraphics[width=0.2\textwidth]{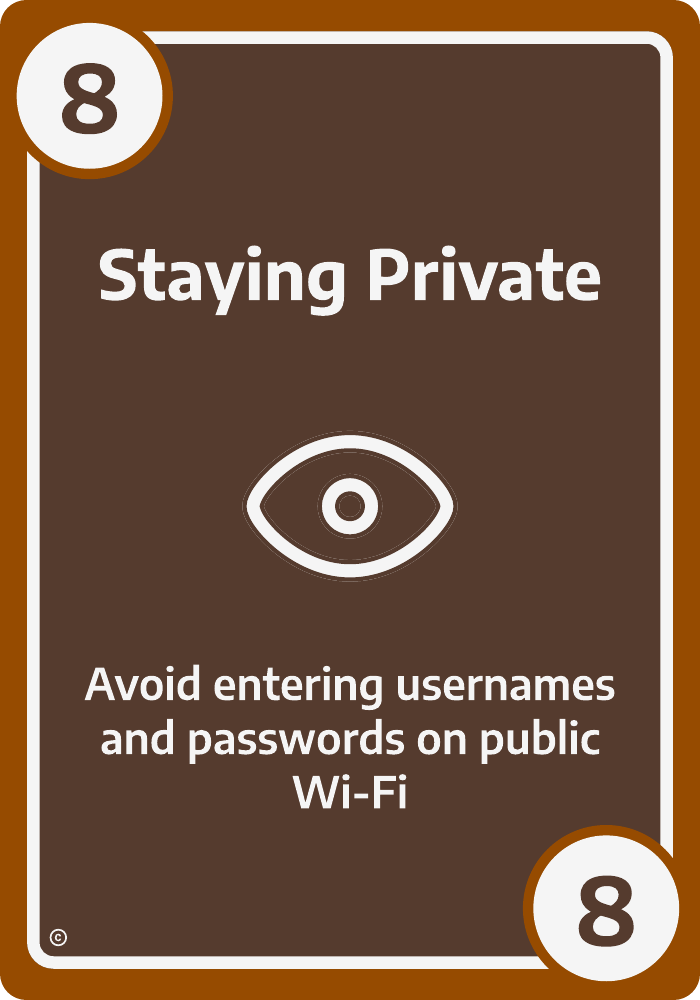}} 
  \subfloat[]{\includegraphics[width=0.2\textwidth]{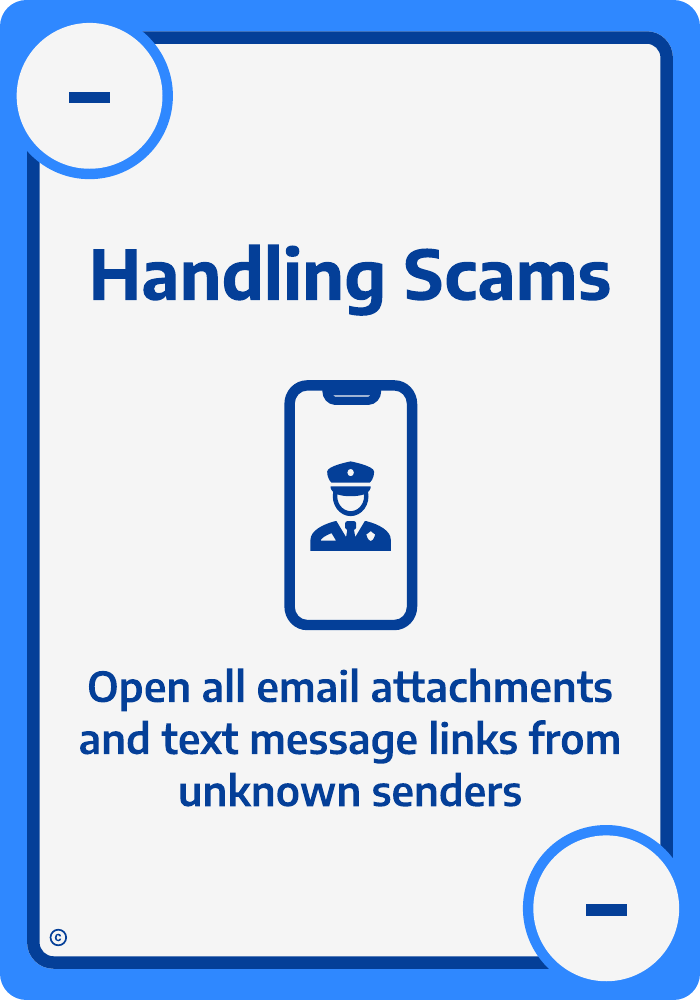}} 
  \subfloat[]{\includegraphics[width=0.2\textwidth]{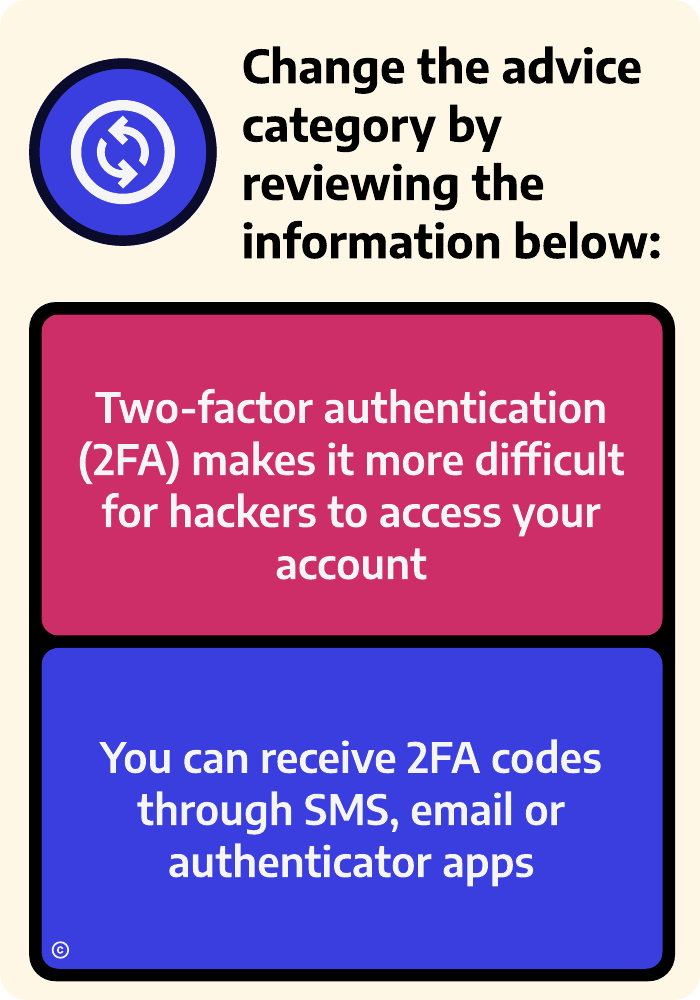}}
  \caption{Examples of each card category from the cybersafety card game.}
  
  \label{fig:cards}
\end{figure*}

\section{Introduction \& Related Work}
Cybersafety remains a core issue for Irish individuals navigating digital spaces. In 2022 alone, cybercrime cost Ireland €10 billion \cite{grantThornton}. Moreover, 50\% of Irish individuals identified cyber threats as the top disaster risk, placing Ireland fourth highest among EU nations \cite{EU-stats}. The problem is two-fold: First, the ever-evolving and diverse threat landscape—from the rise of investment and recovery scams \cite{bpfi_fraud_victims_2025} to the 2021 ransomware attack on the Irish healthcare system \cite{hse_cyber_attack_2025}—has contributed to a pervasive sense of cyber fatigue in an increasingly digital society. Second, while support systems continue to address digital skills training, cybersecurity guidance remains unclear and unreliable ~\cite{murray2023costs, murray2017evaluating, redmiles2020comprehensive}. This can leave citizens feeling vulnerable and uncertain about how to protect themselves, discouraging online participation and deepening digital divides. 

Older adults are particularly at risk, a reality that became especially evident during the COVID-19 pandemic ~\cite{seifert2021double,barnils2022grey,lythreatis2022digital,ramsetty2020impact,tomczyk2022digital}.
This age group faces heightened exposure to privacy and security risks as a result of misconceptions and usability issues \cite{frik2019privacy}, along with the physical and cognitive decline associated with aging ~\cite{schlomann2022older,gitlow2014technology, bhattacharjee2020older,butt2023barriers}. By 2030, 90\% of government services in Ireland are expected to move online \cite{RTE}. Older adults may struggle to access essential services such as telehealth, tax payments or transportation bookings without sufficient cybersecurity awareness. 

To address this issue, cybersecurity experts have proposed game-based learning as a promising method for older adults to learn cyber-safe practices ~\cite{thompson2022gamification}. Game-based elements, such as social learning, incentives and game flow, allow learners to challenge their cybersecurity misconceptions in various contexts to better adapt to targeted vulnerabilities~\cite{dixon2019engaging,chen2019self,yuan2023redcapes,coenraad2020experiencing, ganesh2023tailoring}. Researchers and practitioners mostly lean on digital formats to deliver cybersecurity education, allowing them to provide immediate feedback, customize difficulty levels, and design engaging narratives \cite{zhang2021systematic}. McLennan et al. \cite{mclennan2017evaluation} tested the effectiveness and user-friendliness of a mnemonic approach to password security, using game positions in chess and Monopoly as passwords. The authors reported significant password recall alongside positive user experiences for creating game-based passwords. Gamified elements, such as incremental difficulty levels and tier rankings, adopted by Blackwood-Brown et. al. in their scenario-based application \cite{blackwood2021cybersecurity} were paramount in improving older adults’ knowledge and self-efficacy in cyber threat preventative measures.

Despite the above-mentioned positive learning effects, game-based education in cybersecurity tailored to older adults is rarely touched upon ~\cite{zhang2021systematic}. As demonstrated by Regalado et. al.~\cite{regalado2024game}, ideating educational games which deliver advice on a broad range of cybersecurity domains in digital format is challenging for two reasons: First, older adults might lack basic digital competencies, making it difficult for them to engage with unfamiliar game interfaces. Second, designing games that are both immersive and educationally rewarding for this demographic requires a deep know-how of their specific needs, preferences and cognitive abilities. 

One popular method for delivering cybersafety game-based learning is through playing cards~\cite{zhang2021systematic}. Educational card games fall under the category of `serious games', designed for a primary purpose beyond pure entertainment. Tabletop card games are commonly proposed as effective methods for employees and younger populations to easily infer links between cyber attacks, vulnerabilities and defenses \cite{shah2023cyber,hart2020riskio,priya2022cyber,shah2023introducing}. What sets them apart from regular games is their constructivist learning and training component, allowing players to actively practice specific skills while maintaining the fun and engagement elements of traditional games \cite{kordaki2015constructivist}.  
In a study on non-digital games played by older adults involving 886 participants, playing cards was the most popular, with the benefits of being immersive and cognitively inducing \cite{mortenson2017non,estrada2021cognitive,indarwati2019playing,cardona2023meaningful}. This is observed in common applications of non-digital educational card games, and their constructivist learning effects, for older adults in domains spanning healthy lifestyle adherence \cite{ischak2024analysis}, financial education \cite{chacon2023impact}, human rights awareness \cite{fernandes2023ark}, and intergenerational learning \cite{yuan2024developing}. Continuing this line of research, we posit that older adults benefit from learning cyber-safe practices in low-stakes and non-digital environments, such as those provided through educational card games. To date, only Chung and Yeung \cite{chung2023reducing} have successfully evaluated an educational board game designed for older adults which simulates scam scenarios on a game board and preventative measures as playing cards. Their pre-test/post-test experimental design revealed a significant acquisition of cyber-safe knowledge and self-efficacy in handling scams. 

Our goal was to deliver cybersafety advice to older adults in a social, immersive, and easy-to-understand manner. To achieve this, we propose a cybersafety card game designed not only to engage older adults but also to support digital educators in delivering effective cybersecurity education. 
Although Chung and Yeung \cite{chung2023reducing} provide valuable insights into game-based cybersafety education for older adults, their game
focused exclusively on anti-scam awareness, whereas the current cyber threat landscape is far more diverse. Furthermore, the educational game relies on a game master whose primary role is to oversee the smooth running of gameplay and attainment of key cybersafety practices. Having trained staff as game moderators can assist older adults in making connections between cybersafety topics. However, this approach may not fully engage participants throughout the activity, potentially making the game feel more like a lecture than an interactive experience. Our game takes a different approach: we involve the digital educator as an active player. This allows them to guide the game's flow and explain the rules while also fully participating, enhancing the interactive and immersive nature of the experience.

In our mixed-method protocol, we tested the game among older adults and digital educators in educational settings and sought to answer the following research questions:

\textbf{RQ1:} How did digital educators and older adults perceive the educational cybersafety card game in terms of engagement and enjoyment?

\textbf{RQ2:} To what extent did digital educators and older adults find the cybersafety advice provided through the game useful?

\section{Cybersafety Card Game Design and Instructions}
We developed a novel shedding-type card game consisting of 48 playing cards printed on A6 double-sided polyester material to ensure readability and ease of handling for older adults:

\begin{itemize}
    \item 40 cards are divided into four color-coded advice categories: handling scams (Figure \ref{fig:cards}a), password management (Figure \ref{fig:cards}b), responding to cyber attacks (Figure \ref{fig:cards}c), and staying private (Figure \ref{fig:cards}d). Of these, 32 are ranked from 1 (good advice) to 8 (best advice), whereas 8 `minus' cards represent misconceptions or bad advice for each advice category (Figure \ref{fig:cards}e).
    \item Eight cards in the deck are special `change' cards that enable players to switch between different advice categories (Figure \ref{fig:cards}f).
\end{itemize}

The game’s unique contribution to cybersafety education extends beyond basic cybersafety awareness. It builds on the call for delivering region-specific cybersafety practices \cite{herbert2023world} by highlighting cybercrime preventative measures, reporting mechanisms, and support systems for cybercrime victims tailored to the Irish context (e.g. checkmylink.ie\footnote{\url{https://check.cyberskills.ie/home} is a site safety checker to verify the legitimacy of a website.}, PULSE ID\footnote{PULSE number is a computer-generated identifier for a crime or incident, used by the Gardaí to track the case.}, the Gardaí\footnote{The Gardaí is the national police and security service of Ireland.} and the Crimes Victim Helpline\footnote{The Crime Victims Helpline is a national support service for victims of crime in Ireland; \url{https://www.crimevictimshelpline.ie/}}). 

The cybersafety advice on the cards was developed iteratively through a multi-phase process. 
We began by speaking to Irish digital educators to gain their perspective in digitally up-skilling older adults \cite{gruben2025s}.
During the generative phase of the research, recommendations were crafted in line with best practices from the National Institute of Standards and Technology (NIST) \cite{temoshok2024digital}, EU Frameworks \cite{gdpr_cookies_2025}, and Irish law enforcement \cite{garda_investigation_process_2025}.
The process was further informed by insights gathered from academic research in cybersecurity \cite{zou2024cross, busse2019replication,murray2023costs, sheil2022fianan,murray2017evaluating}. 
The advice was then simplified and validated through focus groups with older adults~\cite{sheil2024enhancing}. 

Feedback from these sessions led to further refinement, incorporating additional valuable advice suggested by participants. The finalized advice can be seen in Table \ref{finaladvice} in Appendix \ref{Advice}.

Players are provided with a 3-page instructional manual (Appendix \ref{Instructions}) that combines text and visuals to explain the gameplay sequentially. To win this game, players must be the first to get rid of all their cards, achieved by playing higher-numbered cards of the same color, cards of the same number from different categories, or change cards. The game begins by shuffling the cards, dealing 7 cards to each player and placing the remaining cards in the draw pile. The first player in the group starts the game by playing one or more numbered cards in any advice category, beginning with a card numbered 1 or higher. Cards are placed in ascending order, if multiple cards from the same category are played. Play proceeds in a clockwise direction, with players taking turns to make their moves.
During their turn, players play cards of the same color as the previous card placed, but with higher numbers, or play cards of the same number from different categories. If they are unable to make a valid move, they have the option to play a change card to switch to a different advice category. If no other options are available, players can play a minus card from the same category but are penalized by taking 2 cards from the draw pile. If none of these moves are valid, then 2 cards should be taken from the draw pile nonetheless.
Special instructions were provided for playing change cards by inferring links between the cybersafety information presented on the cards and the four advice categories.
Participants were encouraged to explain their choice of category when playing the change card, either by explaining a fact they have learned, a personal experience, or their unique cybersecurity strategy. Players were also advised to review the other cards in their hand to help inform their choice when selecting a new advice category. The information presented on the change cards was generated in a way that avoided leading language. For example, a change card presenting phishing-related information did not explicitly include the term `scam', making gameplay more engaging and thought-provoking. 

\section{Evaluation}

This research was approved by our university’s institutional
review board MTU-HREC-FER-24-009-A. Before participation in the study, demographic information was collected using a demographics sheet, as well as providing information sheets and consent forms to ensure informed consent. Participants were also informed that they could play the game regardless of their participation in the study. To acknowledge the value of their time and contribution, each participant received a €40 One4All\footnote{Irish gift voucher; \url{https://www.one4all.ie/}} gift voucher. 

\subsection{Participants}

A convenience sample of 16 participants (14 older adults and 2 digital educators) was approached from two venues in Ireland: St. Goban’s Further Education and Training Board (ETB) in Bantry, Co Cork,  which hosted 7 participants, and the Artane Coolock Family Resource Center in Dublin, hosting 9 participants.
Among the older adults cohort (mean age = 76 years, 13 females), 13 were high-tech users (regularly utilizing various technologies and actively engaging in online activities such as banking and shopping), while 1 participant was classified as a low-tech user (very limited engagement with devices and online activities). Most older adults in our study reported moderate familiarity with cybersafety topics. The digital educators had more than 5 years of experience in their roles, with familiarity with cybersafety topics ranging from moderately familiar to very familiar. Tech usage, digital educator experience, and familiarity with cybersafety topics were rated using 5-point Likert scales included in our demographics sheet.

\subsection{Protocol}

A mixed-method approach was employed, combining qualitative and quantitative methods to evaluate the perceived usefulness of cybersafety advice presented during the card game and participants’ gameplay experience. A concurrent `think-aloud' protocol was used during playtesting, wherein participants were encouraged to verbalize their thoughts, actions and reasoning as they interacted with the game.
Participants were instructed to perform two usability tasks: In the first task, digital educators spent 15 minutes reading through the game instructions and reviewing the deck of cards. In the second task, digital educators briefed older adults about the game and played in tandem in a session lasting approximately one hour. Following playtesting, semi-structured interviews and surveys were conducted to gather additional feedback and perspectives.

Field notes were collected following best practices outlined by Phillippi and Lauderdale \cite{phillippi2018guide}. These notes documented the contextual environment of each session, including the physical setup, player interactions, and non-verbal behaviors such as gestures, expressions and levels of immersion. Detailed observations captured the group dynamics, instances of collaboration or hesitation, and any challenges encountered with the game’s instructions or content. The research team also included reflective notes to identify potential biases, assess their role during the sessions and refine the game’s design iteratively. In tandem with interview and survey data, these notes enriched the analysis by preserving the nuanced details of participant experiences and informing modifications to enhance the game’s accessibility and relevance for older adults and digital educators.

\section{Findings}

\begin{figure*}[h] 
    \centering
    \frame{\includegraphics[width=0.8\textwidth]{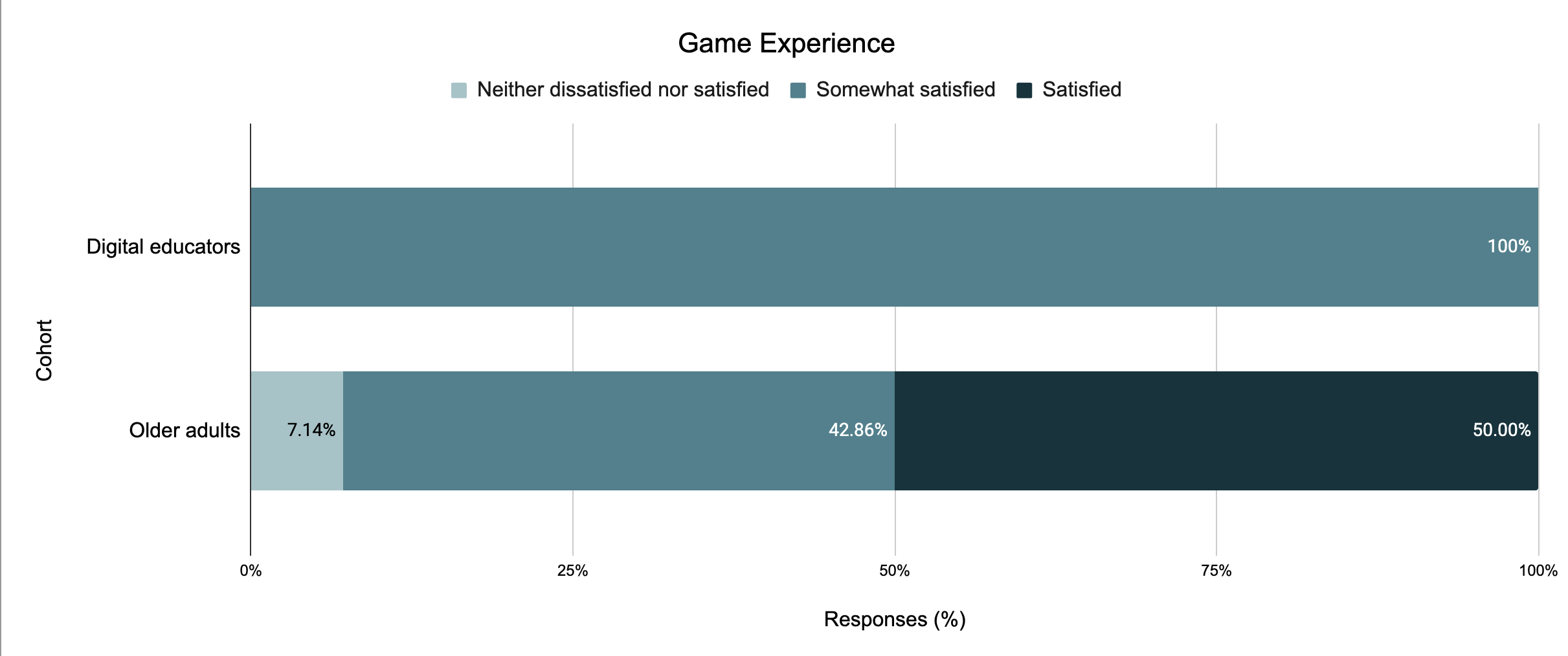}} 
    \caption{Distribution of 5-point Likert scores for game experience (1 = dissatisfied; 5 = satisfied) across digital educators and older adults.}
    \label{fig:scores}
\end{figure*}

\begin{figure*}[h] 
    \centering
    \frame{\includegraphics[width=1\textwidth]{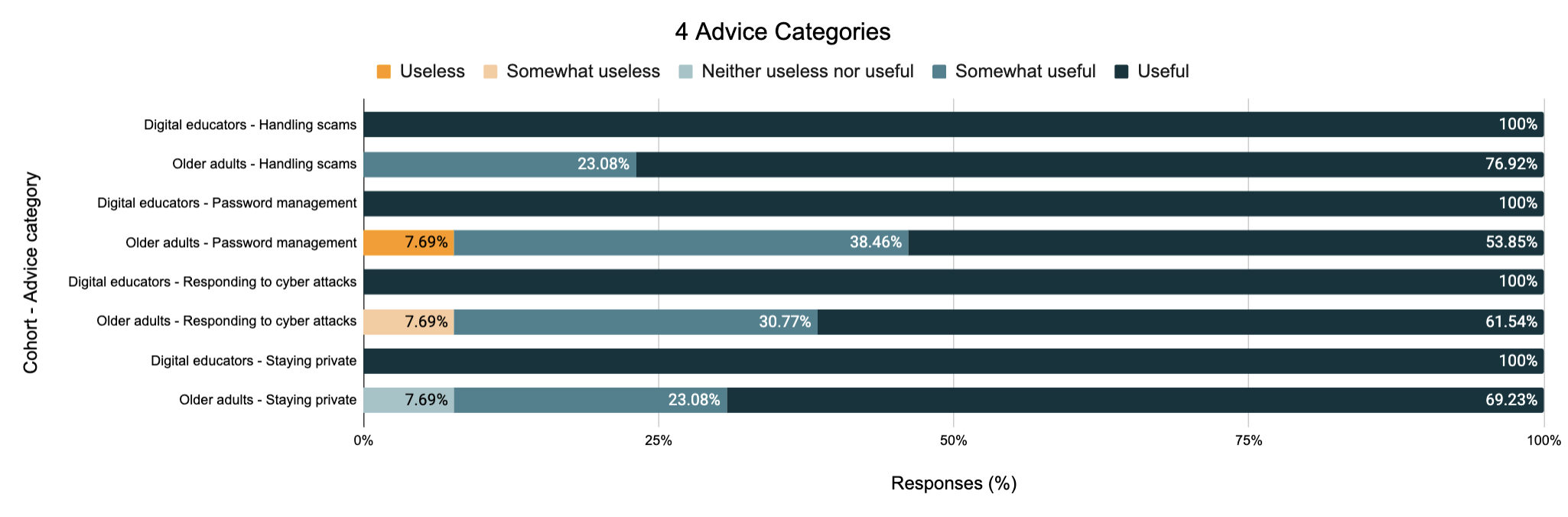}} 
    \caption{Distribution of 5-point Likert scores for the 4 advice categories (1 = useless; 5 = useful) across digital educators and older adults.}
    \label{fig:advicescores}
\end{figure*}

Figures \ref{fig:scores} and \ref{fig:advicescores} present the score distributions for satisfaction with gameplay and the perceived usefulness of the four color-coded advice categories, as collected from post-game surveys. The scores were favorable; however, notable differences emerged between the two demographics. Older adults were more satisfied with the game overall compared to digital educators. Conversely, digital educators rated the advice categories more positively - a finding that will be discussed in Section \ref{discussion}. 

\textbf{RQ1:} How did digital educators and older adults perceive the educational cybersafety card game in terms of engagement and enjoyment?

Participants initially found the game complex and required significant facilitation to understand the rules, card categories and gameplay mechanics. For instance, digital educators commented that the instructions were \textit{``too lengthy''} (P1) and suggested that a video tutorial or prior distribution of the rules could improve the game experience. Older adults expressed the same sentiment: \textit{``Lots of different types of cards..most card games are difficult at first''} (P4).
Engagement improved significantly as participants became more familiar with the game mechanics. Comments like \textit{``I want to play again!''} (P6) and \textit{``You’re making us all interested now''} (P13) reflect growing enjoyment and interest. Additionally, players noted that the game promoted collaborative problem-solving when strategically playing special cards, despite initial struggles.

The Bantry session evolved into a co-creation workshop as participants discussed ways to modify the rules to enhance the learning experience. Integrating advice into gameplay proved difficult, as participants were initially focused on learning the mechanics, overshadowing advice learning. Participants proposed playing a combination move of special cards and color-coded cards which they found are relevant to the change card, allowing them to eliminate most of the cards from their hand easily. The revised rules resulted in participants better engaging with the advice and linking it to their own experiences. Digital educators and older adults suggested making adjustments to the physical cards, some of which include
placing the text further up the cards so that they can easily read them and eliminating the red and orange colors as they had trouble differentiating between the two. 
We identified two key gameplay issues which impeded learning effects: (1) the minus cards contributed little to the overall gameplay, and (2) players often focused excessively on numbers and colors, sometimes neglecting the text.

\textbf{RQ2:} To what extent did digital educators and older adults find the cybersafety advice provided through the game useful?

Participants found the advice on the cards to be highly relevant and valuable. Usefulness scores were the highest for handling scams among older adults, followed by staying private, responding to cyber attacks and password management. Qualitative feedback indicated that the card game provided new knowledge, especially on terms such as 2FA, passphrases and disposable virtual cards. The digital educators appreciated the thought-provoking nature of the cards. For example, the digital educator from Bantry had never encountered the term `PULSE ID'\footnote{PULSE number is a computer-generated identifier for a crime or incident, used by the Gardaí to track the case.} before but found the advice highly useful. Despite this, many admitted that understanding the advice came only after playing multiple rounds, suggesting that the game’s educational impact deepens with repeated play. 

Participants noted the advice was succinct enough for older adults to understand but expressed interest in expanding the advice categories. Online shopping emerged as a prominent topic for both digital educators and older adults, while recognizing scams and applying the advice in practice were priorities for older adults. Older adults highlighted the potential of a non-digital game like this to enhance their digital knowledge, with P14 commenting, \textit{``It’s great for our age group to be learning again''}. Digital educators also recognized the game’s value in sparking meaningful discussions about cybersafety topics. 
 
\FloatBarrier
\begin{figure*}[h]
  \centering 
  \subfloat[]{\includegraphics[width=0.2\textwidth]{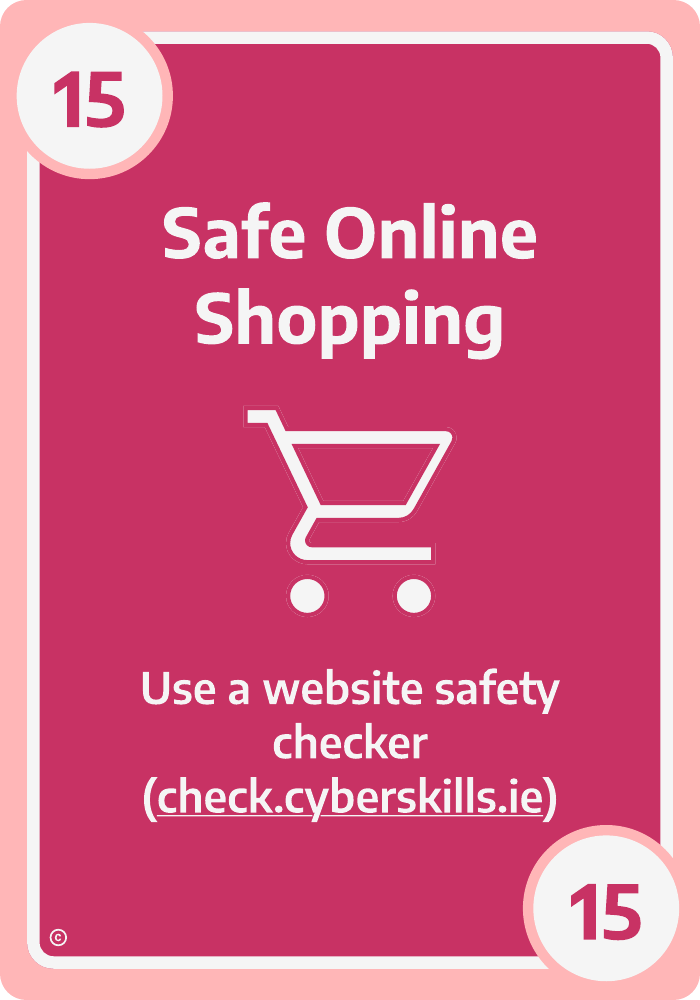}}
  \subfloat[]{\includegraphics[width=0.2\textwidth]{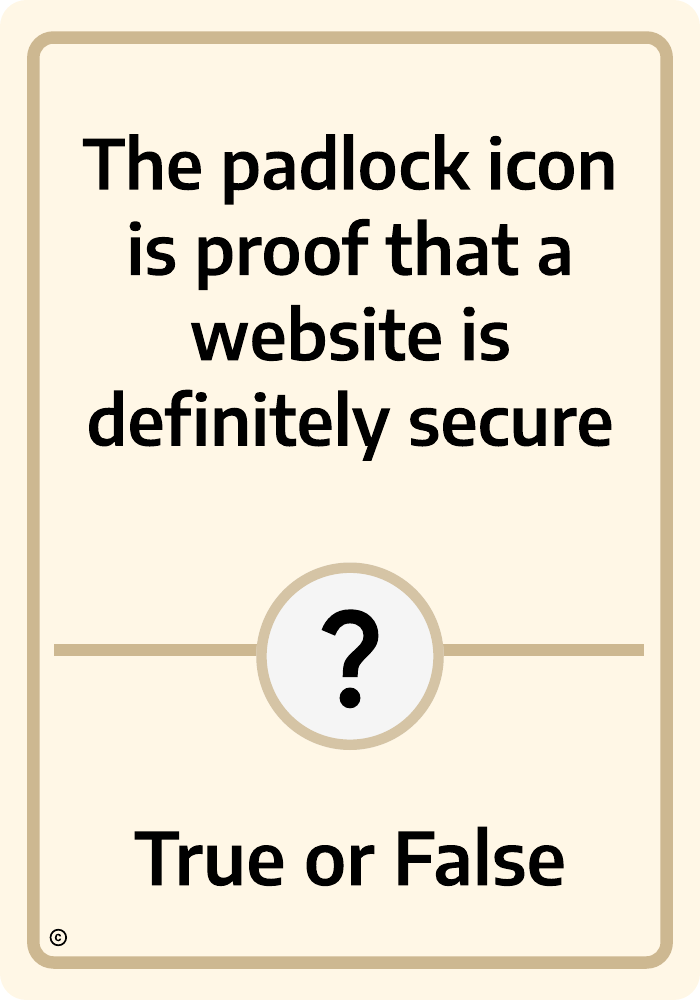}} 
  \subfloat[]{\includegraphics[width=0.2\textwidth]{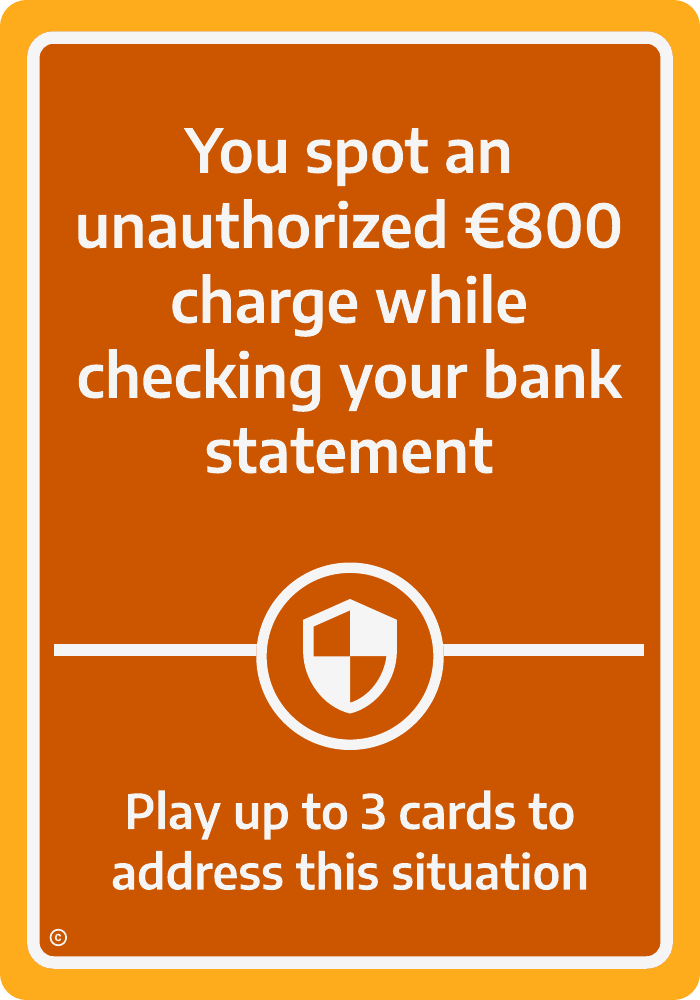}}
  \caption{Modifications to the current card game} 
  \label{fig:2nditeration}
\end{figure*}

\section{Discussion, Limitations and Future Work}
\label{discussion}
The current study has presented the results of a playtesting session aimed at evaluating an educational card game for digital educators to teach cyber-safe practices among older adults. To the best of our knowledge, this is the first non-digital card game to address a broad range of cybersafety topics.
Additionally, the game was designed for digital educators to actively participate in the game with older adults, rather than solely moderating as game masters, as was done previously \cite{chung2023reducing}.

The findings from the `think-aloud' observations, semi-structured interviews and survey results indicate that participants initially faced a steep learning curve in understanding the educational card game. Consistent with previous digital game-based approaches in cybersafety education \cite{blackwood2021cybersecurity}, learning a new game concept did not come naturally for both digital educators and older adults, requiring more time to prepare for the educational card game. However, after a few turns, a flow state emerged: digital educators gained confidence in teaching the game mechanics, while older adults were strategically playing cards to win the game. 
As the game progressed, older adults started to engage more actively with the cybersafety advice, particularly relating advice on handling scams to their personal experiences with relentless scam attempts. Some of the technical terms provided in the cybersafety advice categories—such as 2FA, PULSE ID, passphrases and disposable virtual cards—were unfamiliar to both digital educators and older adults. This sparked thought-provoking discussions during gameplay, encouraging participants to reflect on and explore the advice more deeply. 

Overall, the participants regarded the card game as a promising tool for teaching cyber-safe practices, though modifications on the instructional guide and game setup will be implemented. The instructions guide needs to be revised to include a video tutorial, a detailed description of the advice categories and a supplementary booklet with step-by-step guides on escalating and preventing cybercrimes.  
The cybersafety advice also warrants further modification. As requested by participants, online shopping and recognizing scams deserve more depth in the card game, In response, the final version of the game will feature another color-coded category for safe online shopping (Figure \ref{fig:2nditeration}a) and an updated handling scams category incorporating phishing-related information across SMS and email communications. 

As mentioned in findings, we observed that the minus card and lack of attention to the text on the cards hindered the learning experience. To address these issues, the minus cards will be repurposed into `True and False' cards (see Figure \ref{fig:2nditeration}b) representing a mix of cybersecurity misconceptions and facts, which players can answer when unable to play same-colored or same-numbered cards. If players guess incorrectly, they must draw two cards from the draw pile. The latter issue will be addressed by replacing the change cards with `Scenario' cards (Figure \ref{fig:2nditeration}c). Drawing inspiration from Chung and Yeung \cite{chung2023reducing}, these cards will form part of an attack-defense scenario, requiring participants to apply the advice in practice by playing up to three color-coded cards to prevent and escalate cyber attacks. This gives players a chance to discard cards from their hand, but only if the defense situations are relevant to the cyber attack. Both sets of cards will be randomly shuffled and played as a separate `Challenge' pile to enhance engagement. 
The findings also highlighted the observation of one participant who noted the similarity between the orange and red cards. In future iterations, specialized color palettes for individuals with color vision deficiency will be considered using tools such as ColorMaker~\cite{salvi2024color}.

This study is not without its methodological limitations. First, recruiting only two digital educators limits the generalizability of the findings. Moving forward, the cybersafety educational card game should be tested with a representative sample of digital educators across various educational settings. Second, the observed learning effects were confined to a two-hour playtesting session. Future research should explore these effects by adopting a pre and post-test experimental design with an intervention in between, allowing for a longitudinal assessment of the game’s educational impact. Lastly, one participant rated the color-coded password management advice as unhelpful, the reason for this remains unclear, as the survey was anonymous and did not include an open-ended option for follow-up on the Likert scale responses. Future surveys will be designed to enable participants to provide more detailed feedback.

The playtesting sessions described in the present paper have shown that a cybersafety card game can potentially be a useful tool for informal digital educational settings.
While the game remains a work in progress, insights from our playtesting sessions will inform further refinements not only on the gameplay, but also with the cybersafety advice provided.

\begin{acks}
This publication has emanated from research conducted with the financial support of the EU Commission Recovery and Resilience Facility under Research Ireland Our Tech Grant Number 22/NCF/OT/ 11212G. We thank the Irish National Cyber Security Centre for their support in the realization of this work. Dr Murray is supported by Taighde Éireann – Research Ireland under Grant number 13/RC/2077\_P2 at CONNECT: the Research Ireland Centre for Future Networks. This publication was supported in part by a Google Trust and Safety Research Award and the CyberSkills HCI Pillar 3 Project $18364682$.
\end{acks}

\bibliographystyle{ACM-Reference-Format}
\bibliography{Bibliography}

\appendix
\section{Semi-structured Interview Questions}

\textbf{Task 1: Digital educators reading the instructions and reviewing the cards}

\textbf{Q1.} How would you describe your experience of reading through the game instructions? 
    
\begin{itemize}
    \item Prompts: How easy or difficult did you find the instructions? How comfortable or uncomfortable would you feel using these instructions to play the game on your own? What would you change about the instructions, if anything?
\end{itemize}

\noindent
\textbf{Task 2: Digital educators playing the game with older adults. }

\textbf{Q2.} How would you describe your experience of playing the card game?
    
\begin{itemize}
    \item Prompts: What did you enjoy most about playing the game? Do you have any thoughts on the size and the color of the playing cards? Did you find it easy or difficult to understand the text presented on the cards? Where the cards easy or difficult to handle? What would you change about the card game, if anything?
\end{itemize}

\textbf{Q3.} What did you learn about cybersafety advice from playing the game? 
    
\begin{itemize}
    \item Prompts: How easy or difficult was the advice to understand? How comfortable would you feel using the advice on your own, in your everyday life going forward? Was there something missing from the advice that you’d like to learn more about? What would you change about the advice, if anything?
\end{itemize}

\newpage   
\section{Survey Questions}

\textbf{Q1.} How would you rate each of the 4 advice categories (handling scams, password management, responding to cyber attacks, and staying private), from `useless' to `useful'? (5-point Likert scale).

\noindent
\textbf{Q2.} How would you rate your overall experience with the card game, from `dissatisfied' to `satisfied'? (5-point Likert scale).

\noindent
\textbf{Q3.} Which piece of advice from the 4 advice categories or the change cards stood out to you the most during the game, and why? (Open-ended question).

\noindent
\textbf{Q4.} What do you feel was missing from the advice categories provided?
(Open-ended question).

\onecolumn
\section{Cybersafety Advice \& Information}

\label{Advice}

\begin{table*}[htbp]
\centering

\renewcommand{\arraystretch}{1.3} 
\setlength{\tabcolsep}{8pt} 
\footnotesize 
\caption{Cybersafety practices and misconceptions categorized into four areas: Handling Scams, Password Management, Responding to Cyber Attacks, and Staying Private.}
\begin{tabular}{p{3cm} p{3.1cm} p{3.2cm} p{3.1cm}} 
\hline
\textbf{Handling Scams} & \textbf{Password Management} & \textbf{Responding to Cyber Attacks} & \textbf{Staying Private} \\ \hline
\multicolumn{4}{c}{\textbf{Numbered Cards}} \\ \hline
1. Block suspicious calls and texts & 1. Use `remember my password' on personal devices & 1. Stay calm and assess the situation & 1. Only fill in mandatory fields in forms \\ 
2. Be wary of sites found through unsolicited emails or pop-up ads & 2. Never use `remember my password' on shared or public devices & 2. Immediately change your compromised passwords and log out of all devices & 2. Decline unnecessary cookies to avoid tracking online \\ 
3. Be cautious of deals that seem too good to be true & 3. Use a 3-word sentence as your password & 3. Check if other accounts (social media, email) have been compromised (haveibeenpwned.com\footnote{\url{https://haveibeenpwned.com/}}) & 3. Select `manage' or `options' on cookie pop-ups and select `necessary only' \\ 
4. Use Google Search to check if it’s a known scam & 4. Write down your passwords and keep them in a secure location & 4. Enable two-factor authentication (2FA) & 4. Reject unnecessary requests from websites/apps to get your location and contacts \\ 
5. Agree on a `safe word' with family members to verify suspicious communications & 5. Avoid using the same passwords for multiple accounts & 5. Collect evidence (e.g., take screenshots) & 5. Never share personal information when talking to voice-assisted technology \\ 
6. Collect evidence (e.g., take screenshots) if you are a victim of a scam & 6. If your password is compromised, change it immediately & 6. Contact your service provider (e.g., bank) by calling, physically going in, or going online & 6. Limit the amount of personal information shared on social media \\ 
7. Check links before you click (check.cyberskills.ie) & 7. Consider using a digital password manager & 7. Go to your local Garda station and request a PULSE ID & 7. Regularly review and update privacy settings on your devices and accounts \\ 
8. Go directly to the source (e.g., bank) by calling, physically going in, or going online & 8. Enable two-factor authentication (2FA) & 8. Provide the PULSE ID to your service provider so they can rectify the situation & 8. Avoid entering usernames and passwords on public Wi-Fi \\ \hline
\multicolumn{4}{c}{\textbf{Minus Cards}} \\ \hline
- Open all email attachments and text message links from unknown senders & - Use your date of birth and the names of family members as your passwords & - Ignore the hack and continue using your personal accounts & - The padlock icon is proof that a website is definitely secure \\ 
- Scammers only target people for large sums of money & - Always change passwords on a regular basis & - Anti-virus software is enough to keep you protected & - Declining cookies denies you access to the website \\ \hline
\end{tabular}

\label{finaladvice}
\end{table*}

\begin{table*}
\centering
\renewcommand{\arraystretch}{1.3} 
\setlength{\tabcolsep}{12pt} 
\small 
\caption{Cybersafety concepts and best practices presented on the change cards.}
\begin{tabular}{c p{11.5cm}} 
\hline
\textbf{Change Cards} & \textbf{Information} \\ \hline 
\textbf{Card 1} & - Phishing attempts often mimic trusted sources. 

- Cyber con artists use urgency, fear, or curiosity to obtain your personal information. \\ \hline
\textbf{Card 2} & - 2FA makes it more difficult for hackers to access your account. 

- You can receive 2FA codes through SMS, email, or authenticator apps. \\ \hline
\textbf{Card 3} & - The PULSE ID is a unique number allocated to an incident in the Garda System. 

- Expect a Garda Victims Service Office letter with the Garda's name and PULSE ID after you report a cybercrime. \\ \hline
\textbf{Card 4} & - Cookies remember your browsing preferences, location, and login information. 

- Reject unnecessary cookies to keep your personal information safe. \\ \hline
\textbf{Card 5} & - Google Search is an effective method for verifying information across multiple sources. 

- Pay attention to customer reviews and news sources. \\ \hline
\textbf{Card 6} & - Passphrases are a combination of words used to secure access to your accounts. 

- An example of a passphrase would be `I make tea at 9:30am'. \\ \hline
\textbf{Card 7} & - `Compromised', `breached', `hacked', and `unauthorized access' all refer to similar situations. 

- Contact the Crime Victims Helpline at 116006 for support after a cybercrime. \\ \hline
\textbf{Card 8} & - Use a disposable virtual card from a trusted bank to protect your main card when shopping online. 

- Regularly check your transactions for any unauthorized activity. \\ \hline
\end{tabular}

\label{tab:changecards}
\end{table*}

\onecolumn
\section{Card Game Instructions}
\vspace{1.5em}
\noindent
\captionsetup{type=figure}
\centering
\frame{\includegraphics[width=0.7\textwidth]{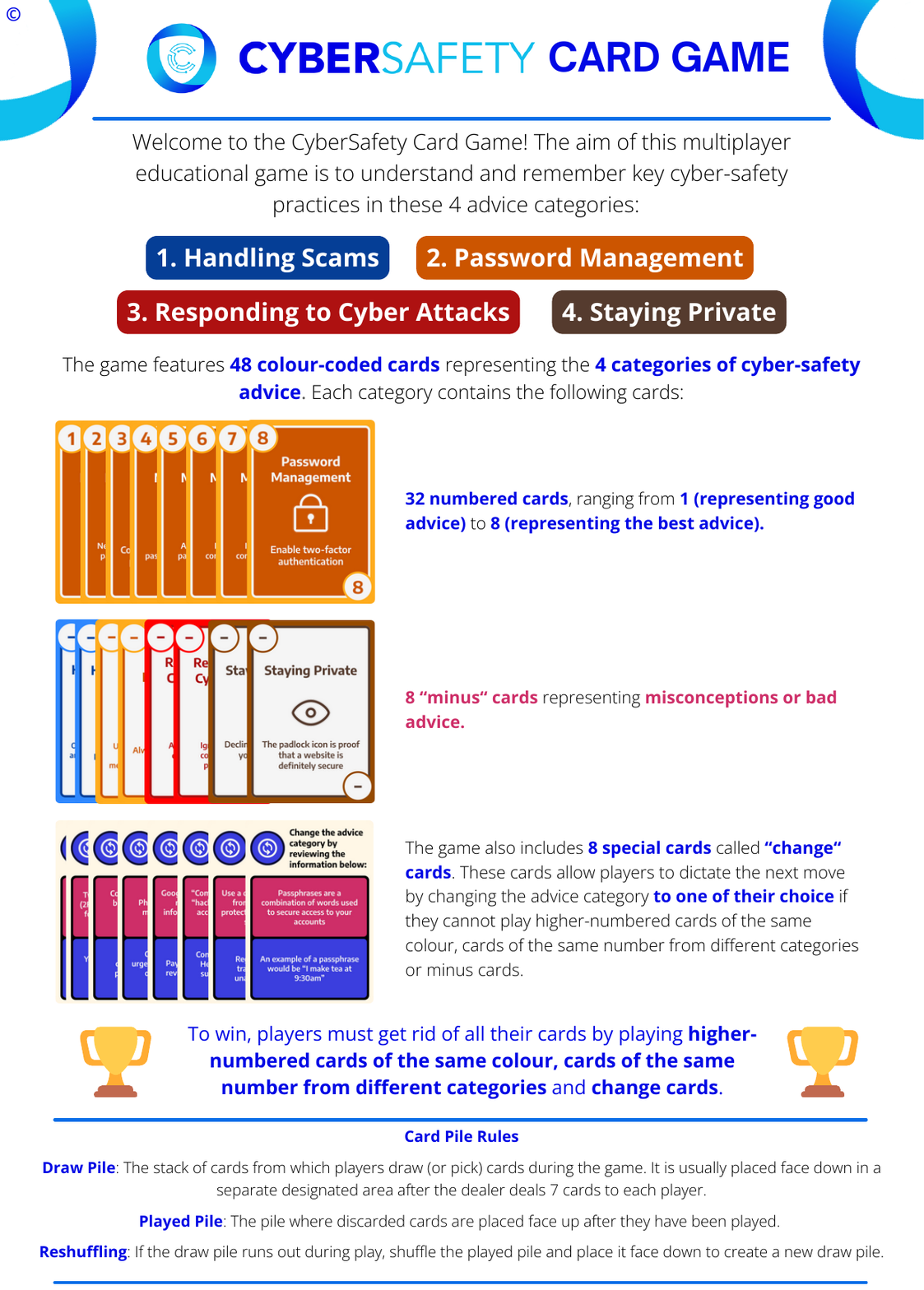}}
\caption{First page of the game instructions.}

\begin{figure} 
 \centering
\frame{\includegraphics[width=0.7\textwidth]{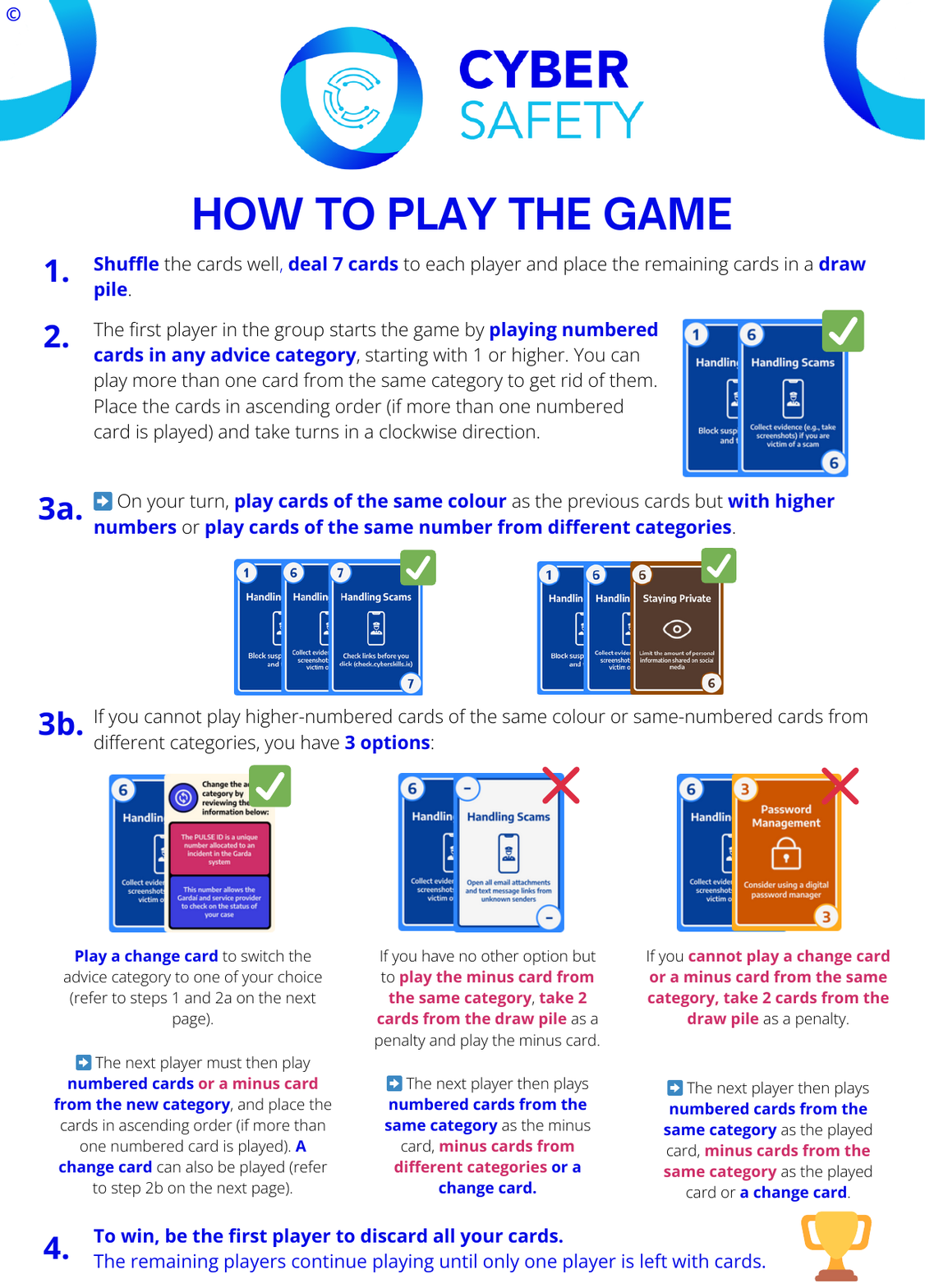}} 
      \caption{Second page of the game instructions.}
    
    \label{fig:page2} 
\end{figure}

\begin{figure} 
    \centering
\frame{\includegraphics[width=0.7\textwidth]{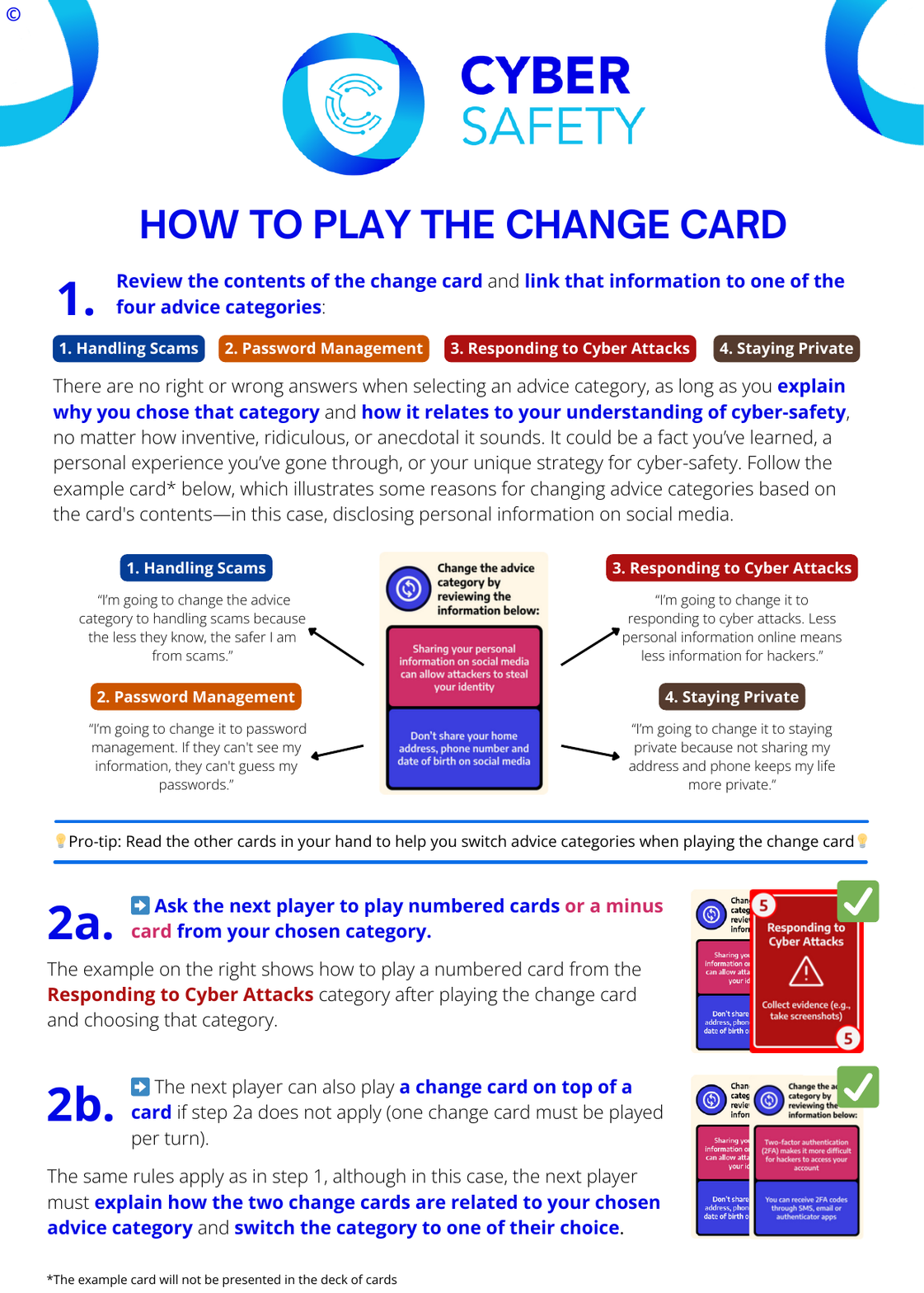}} 
    \caption{Third page of the game instructions.}
  
    \label{fig:page3} 
\end{figure}

\end{document}